\documentclass[fleqn,10pt]{wlscirep}
\usepackage{graphicx}
\usepackage{caption}
\usepackage{subcaption}
\usepackage{times}
\usepackage{bm}
\usepackage{amsmath, amsthm, amssymb}
\DeclareMathOperator*{\argmin}{arg\,min}
\usepackage{color}

\newcommand{\psf}{ps\!f}

\title{3D Snapshot Microscopy of Extended Objects}

\author[1]{Xiang Huang}
\author[2]{Alan Selewa}
\author[2]{Xiaolei Wang}
\author[2]{Matthew K. Daddysman}
\author[2]{Itay Gdor}
\author[3]{Rosemarie Wilton}
\author[3]{Kenneth M. Kemner}
\author[4]{Seunghwan Yoo}
\author[4]{Aggelos K. Katsaggelos}
\author[4]{Kuan He}
\author[4]{Oliver Cossairt}
\author[1]{Nicola J. Ferrier}
\author[1,*]{Mark Hereld}
\author[2,5]{Norbert F. Scherer}

\affil[1]{Division of Mathematics and Computer Science, Argonne National Laboratory, Lemont, IL 60439, USA}
\affil[2]{James Franck Institute, University of Chicago, Chicago, IL 60637, USA}
\affil[3]{Division of Biology, Argonne National Laboratory, Lemont, IL 60439, USA}
\affil[4]{Department of EECS, Northwestern University, Evanston, IL 60208, USA}
\affil[5]{Department of Chemistry, University of Chicago, Chicago, IL 60637, USA}

\affil[*]{corresponding author: hereld@anl.gov}

\begin{abstract}
Volumetric biological imaging often involves compromising high temporal resolution at the expense of high spatial resolution when popular scanning methods are used to capture 3D information.  
We introduce an integrated experimental and image reconstruction method for capturing dynamic 3D fluorescent extended objects as a series of synchronously measured 3D snapshots taken at the frame rate of the imaging camera.
We employ multifocal microscopy (MFM) to simultaneously image at 25 focal planes and process this depth-encoded image to recover the 3D structure of extended objects, such as bacteria, using a sparsity-based reconstruction approach.
The combined experimental and computational method produces image quality similar to confocal microscopy in a fraction of the acquisition time.  
In addition, our computational image reconstruction approach allows a simplified MFM optical design by correcting aberrations using the measured response to point sources. 
This ``compressive" MFM acquisition and reconstruction method, where an image volume with roughly 8 million voxels is recovered from a single 1-megapixel captured image, enables straightforward study of dynamic processes in 3D, and as a simultaneous snapshot advances the state of the art in dynamic 3D microscopy.
\end{abstract}

\begin{document}

\flushbottom
\maketitle
\thispagestyle{empty}

\section*{Introduction}

Modern imaging systems such as spinning-disk confocal~\cite{minsky1961microscopy,Nakano2002} and light-sheet microscopy~\cite{Voie1993,Chen2014} can capture high-fidelity 3D images by sequential refocusing, where the objective captures a single focal plane at a time as it scans through the sample, typically requiring seconds to acquire the entire target volume~\cite{wang2005performance}. 
In a cell, however, typical time scales of interest are on the order of milliseconds, which cannot be resolved with sequential refocusing~\cite{Tabei2013,Barlan2013}.

Multifocal microscopy~\cite{Abrahamsson2012}, an advance over multifocal plane microscopy (MUM)~\cite{Prabhat2004,Ram2008}, provides a way to capture 3D data in a single synchronous snapshot, thereby removing the limitations of sequential refocusing.  
The potential of multifocal imaging for obtaining volumetric information can be appreciated by realizing that an out-of-focus image of a point source provides better localization of its z-position than an in focus image~\cite{Prabhat2004}. 
Our goal is to reconstruct 3D information about biological objects from MFM images.
Previous MFM systems~\cite{Ram2008,Abrahamsson2015,Oudjedi2016} have developed approaches that localize \emph{point} sources.
These approaches will not be applicable to recovering extended objects, such as unresolved distributions of fluorescent molecules with a bacterium.
Here we present an approach that will recover the detailed distribution of fluorescence in an extended object.

\section*{Results}

\begin{figure}[htb]
\begin{subfigure}{1.0\linewidth}
\includegraphics[width=1\linewidth]{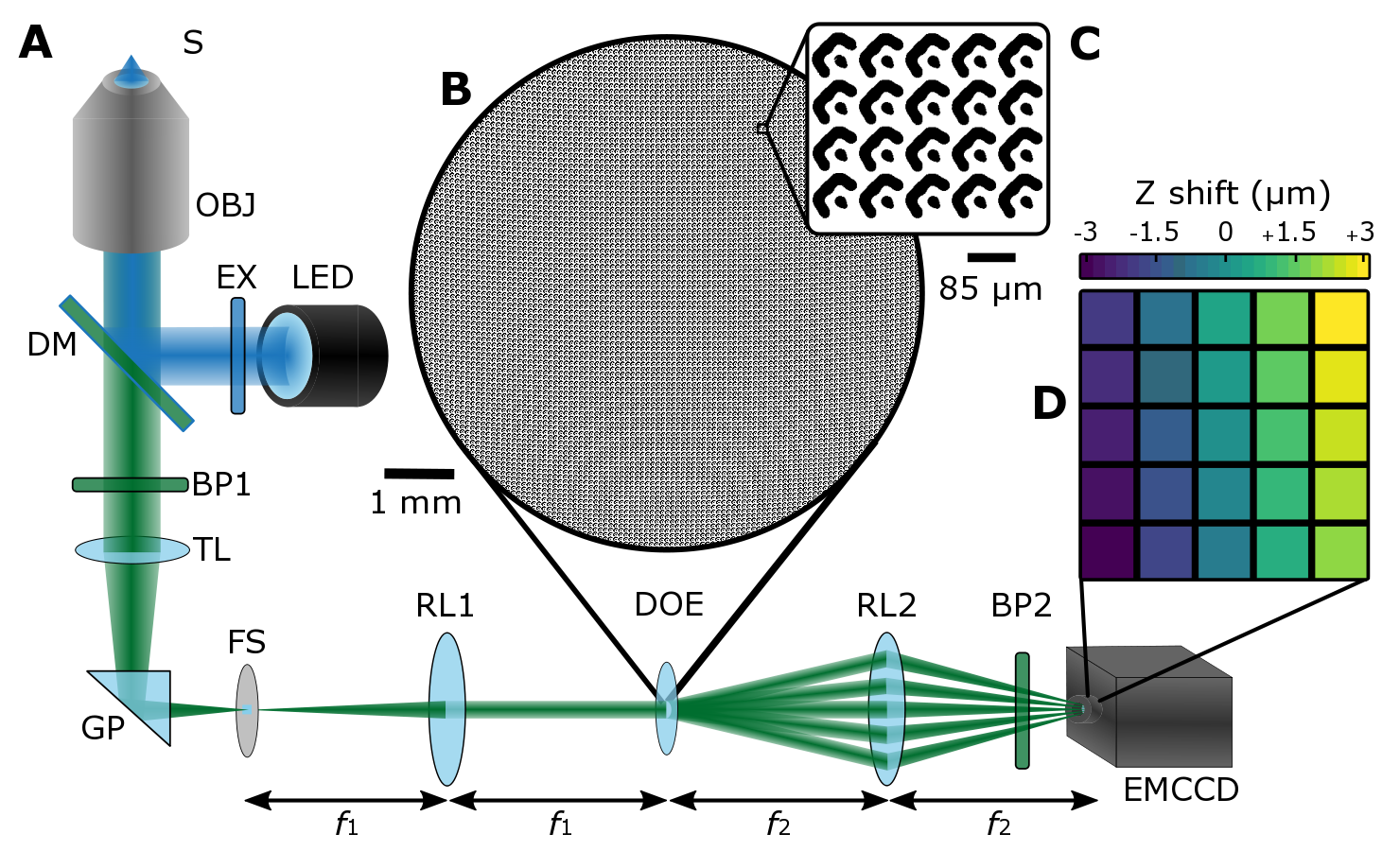}
\subcaption{Schematic of 3D snapshot MFM microscope.}
\end{subfigure}

\vspace{0.2in}
\begin{subfigure}{1.0\linewidth}
\includegraphics[width=0.9\linewidth]{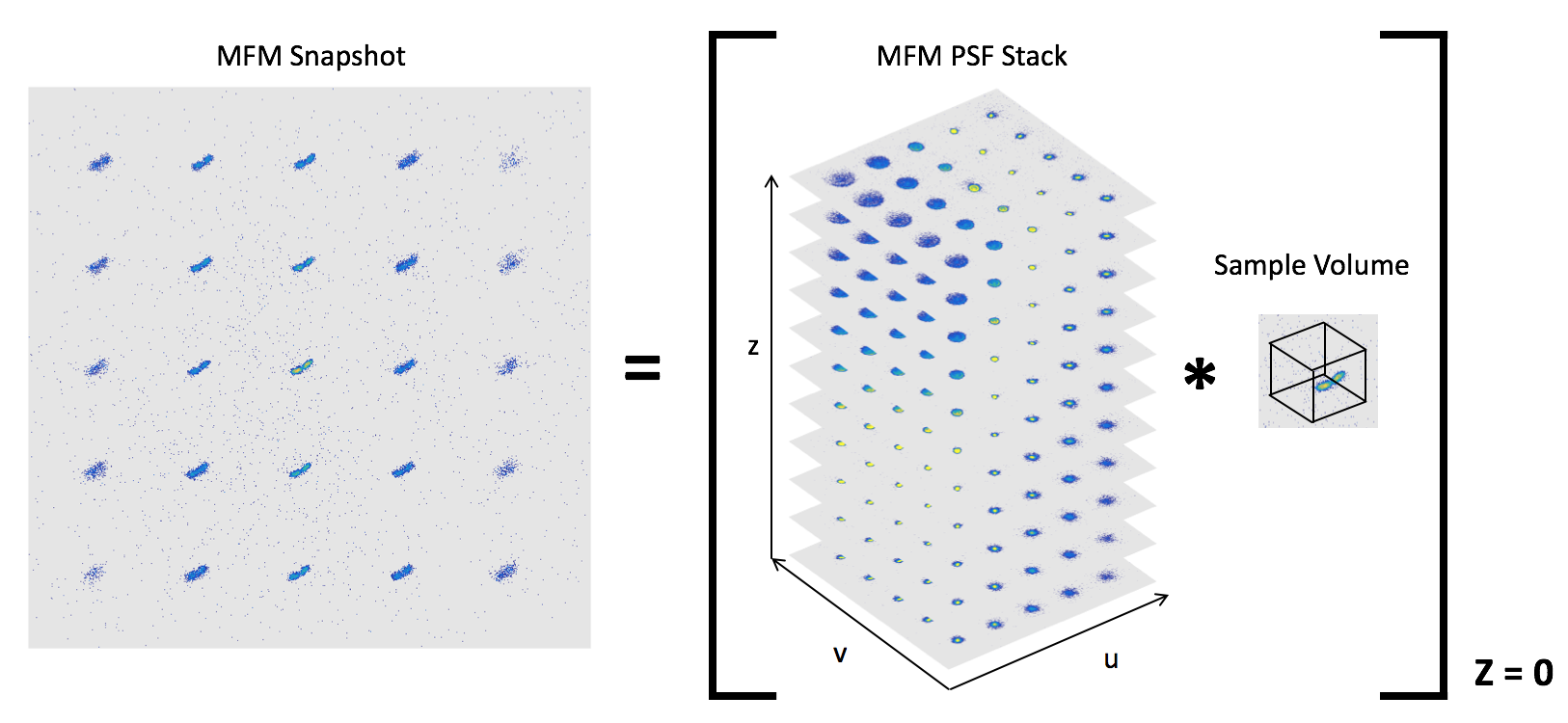}
\subcaption{The forward model maps the 3D sample volume to a 2D MFM image.}
\end{subfigure}

\caption{Overview of the 3D snapshot MFM microscope.  (a) The image of the sample volume (S) at the 1~mm$^2$ field stop (FS) is relayed through a 4-f system to the detector (EMCCD). The diffractive element produces 25 focal shifted sub-images at 250~nm intervals of the sample volume on the detector.
(b) Schematic diagram of the PSF showing MFM snapshots of a point source comprising an image stack. The left hand side shows an MFM snapshot consisting of 5x5 depth-encoded image ``tiles" that results from the convolution of the sample volume with the PSF.}
\label{fig:MFMinstrument}
\end{figure}

We introduce a 3D computational imaging method that overcomes these limitations and demonstrate it on dynamical extended objects by capturing the volumetric data required for reconstruction at 25 Hz (limited only by the detector frame rate). As shown in figure~\ref{fig:MFMinstrument}, the approach combines (i) capture of depth-encoded snapshot acquired images based on multifocal microscopy (MFM), and (ii) a method for sparse reconstruction of the 3D volume from the 2D snapshot.

\subsection*{Optical Setup}
For MFM the light from an infinity-corrected objective is split into several beams where each beam can be understood to create an image with a tube lens with a different focal length. Therefore, each tube lens-objective combination provides a widefield image of a different focal plane in the sample volume. The speed of 3D acquisition is limited only by the camera frame rate and the brightness of the sample. 
Abrahamsson et al.~\cite{Abrahamsson2012} demonstrated MFM using a novel diffractive optical element and a single detector, a configuration we adapt here (figure~\ref{fig:MFMinstrument}(a)). 
Each lens-objective combination forms a spatially distinct sub-image at the detector plane. 

A single diffractive element (DOE) provides depth-encoding of our 3D sample volume. We designed the diffractive element to split the beam into an array of $5 \times 5$ differently focused tiles on the detector following the procedure outlined by Abrahamsson et al.~\cite{Abrahamsson2012} for their multi-focus grating (see Supplementary Materials for details). Our optical design forgoes the corresponding chromatic correction grating and custom multi-faceted prism. The chromatic aberration introduced by the DOE for the 15~nm bandpass for each beam is accounted for through its signature in the PSF.

\subsection*{The Snapshot and PSF}
To inform the 3D reconstruction, we measure the response of the system by collecting MFM images of a nominal point source at 50 nm intervals along $z$ (figure~\ref{fig:MFMinstrument}(b)). An individual slice of the PSF includes 25 widefield exposures spanning $z$ in 250~nm focus increments, each ``distorted'' by its own chromatic aberrations from the DOE. 
For extended objects, details about the sample structure are obfuscated by the out-of-focus sub-images, because MFM is a widefield microscopy.

Given a captured MFM image of an object, our algorithm estimates the 3D object image that produces the best match to the measured 2D MFM image using a sparsity-based reconstruction framework~\cite{Baraniuk2007} %~\cite{Beck2009}.
By doing so, the resulting solution incorporates all of the data in the MFM image, using the information captured in the PSF about correlations between pixel intensities in the 25 focal tiles and through the fine grain $z$-slicing of the PSF.

\subsection*{3D Reconstruction}
The algorithm is implemented by discretizing a sample volume, $\bm{f}$, into 108 nm $\times$ 108 nm $\times$ 50 nm voxels, matching the pixel scale and the $z$-step size of the PSF.
For each $z$-plane of the object volume we compute its contribution to the simulated MFM image, $\bm{g}$, by performing a 2D convolution with the corresponding plane in the PSF. 
Summing these contributions gives the forward projection of the object volume through the MFM optical system. 
This prescription of the forward model, $\bm{g} = \bm{Af}$, for MFM imaging is equivalent to extracting the $z = 0$ plane of the volume convolved with the 3D PSF (figure~\ref{fig:MFMinstrument}). It maps a 3D volumetric fluorescence distribution in object space (e.g., from a bacterium) to a 2D MFM image in the microscope's detector space (see supplementary material).

Using this forward imaging model, we formulate the volume reconstruction as an optimization problem~\cite{Baraniuk2007,Candes2006,Donoho2006}. 
The dimension of the 3D volume to be reconstructed is much larger than the dimension of the observed data (i.e., the 2D image on the detector). 
Hence, direct inversion of the forward imaging model, $\bm{g} = \bm{Af}$, to find $\bm{f}$ is an ill-posed problem~\cite{tikhonov}. 
We take a well-studied approach~\cite{Carrington1990,Carrington1995} to this circumstance by instead solving a constrained optimization problem that simultaneously minimizes the error between the fit of data to the forward model and a constraint enforcing a prior model on the form of the solution,
\begin{equation}
\bm{\hat{f}} = \operatorname*{argmin}_{\bm{f} \geq 0} \quad \frac{1}{2}  \left\| \bm{g} - \bm{Af} \right\|^2  + \lambda {J(\bm{f})}. 
\label{eq:argmin}
\end{equation}
The prior model, ${J(\bm{f})}$, can be selected to encode prior knowledge about the solution such as smoothness or sparsity of the signal (see Supplementary Materials for details).  In our results we use the L1 norm of $\bm{f}$, which favors sparsely filled solutions. We use TwIST~\cite{Beck2009} modified to incorporate non-negative thresholding at each iteration to solve the optimization problem in Equation (\ref{eq:argmin}).

\subsection*{Measurements}
\begin{figure*}[ht!]
\begin{minipage}[b]{0.35\linewidth}
\centering
\begin{subfigure}[b]{1\linewidth}
  \centering
  \includegraphics[width=0.87\linewidth]{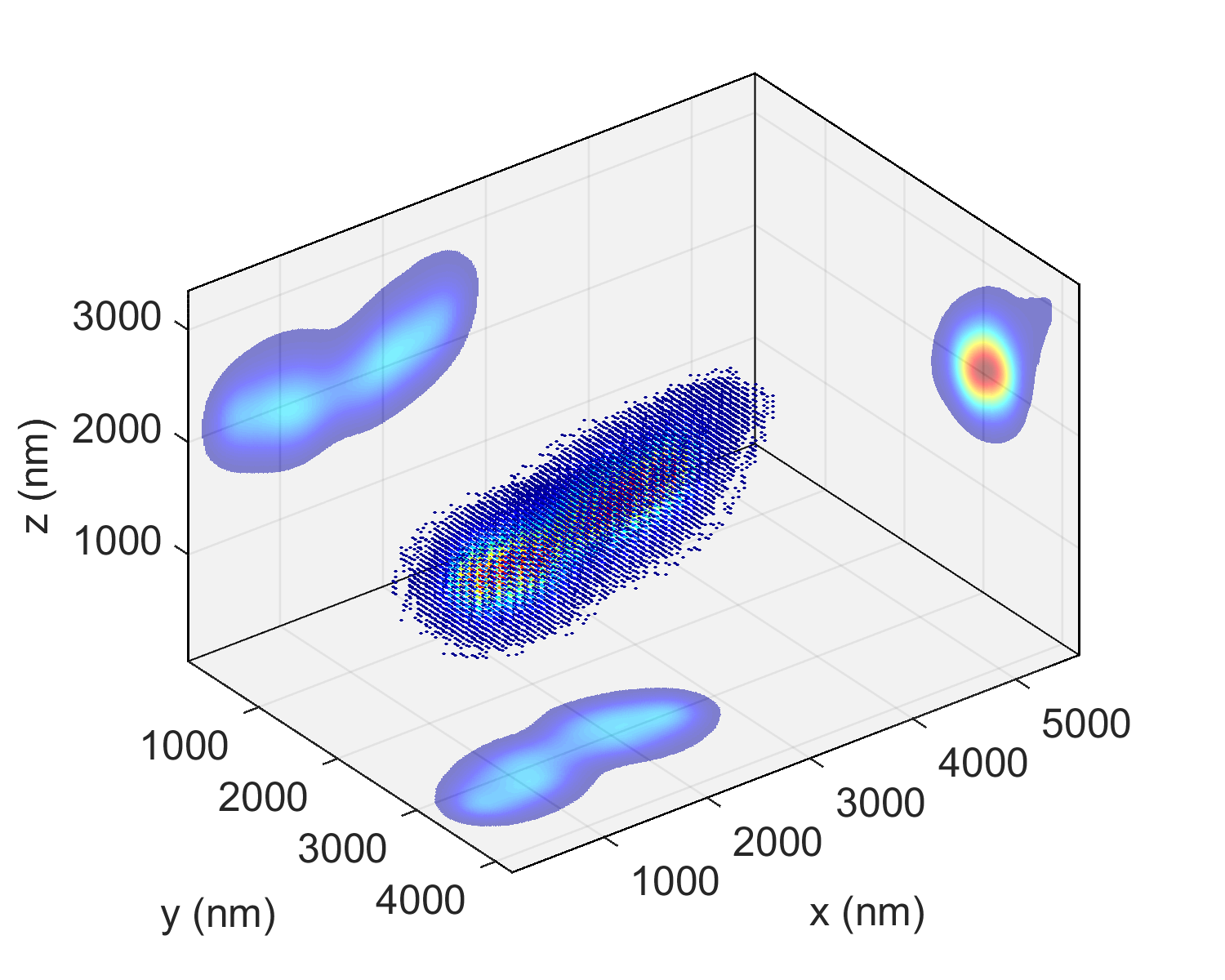}     \hspace{-0.2in}
  \includegraphics[width=0.11\linewidth]{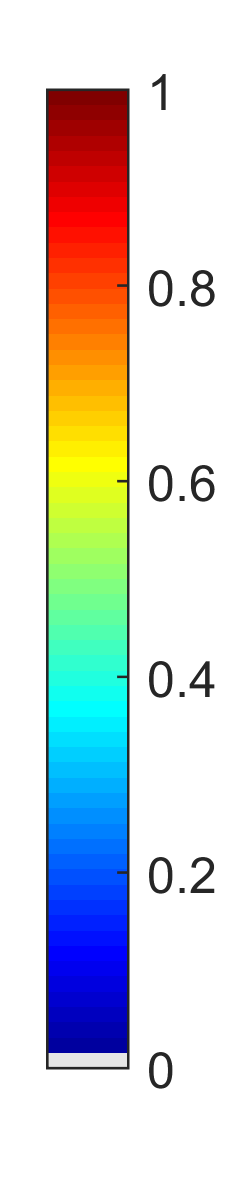}
\subcaption{Confocal $z$-stack} 
\end{subfigure}

\begin{subfigure}[b]{1\linewidth}
  \centering
  \includegraphics[width=0.87\linewidth]{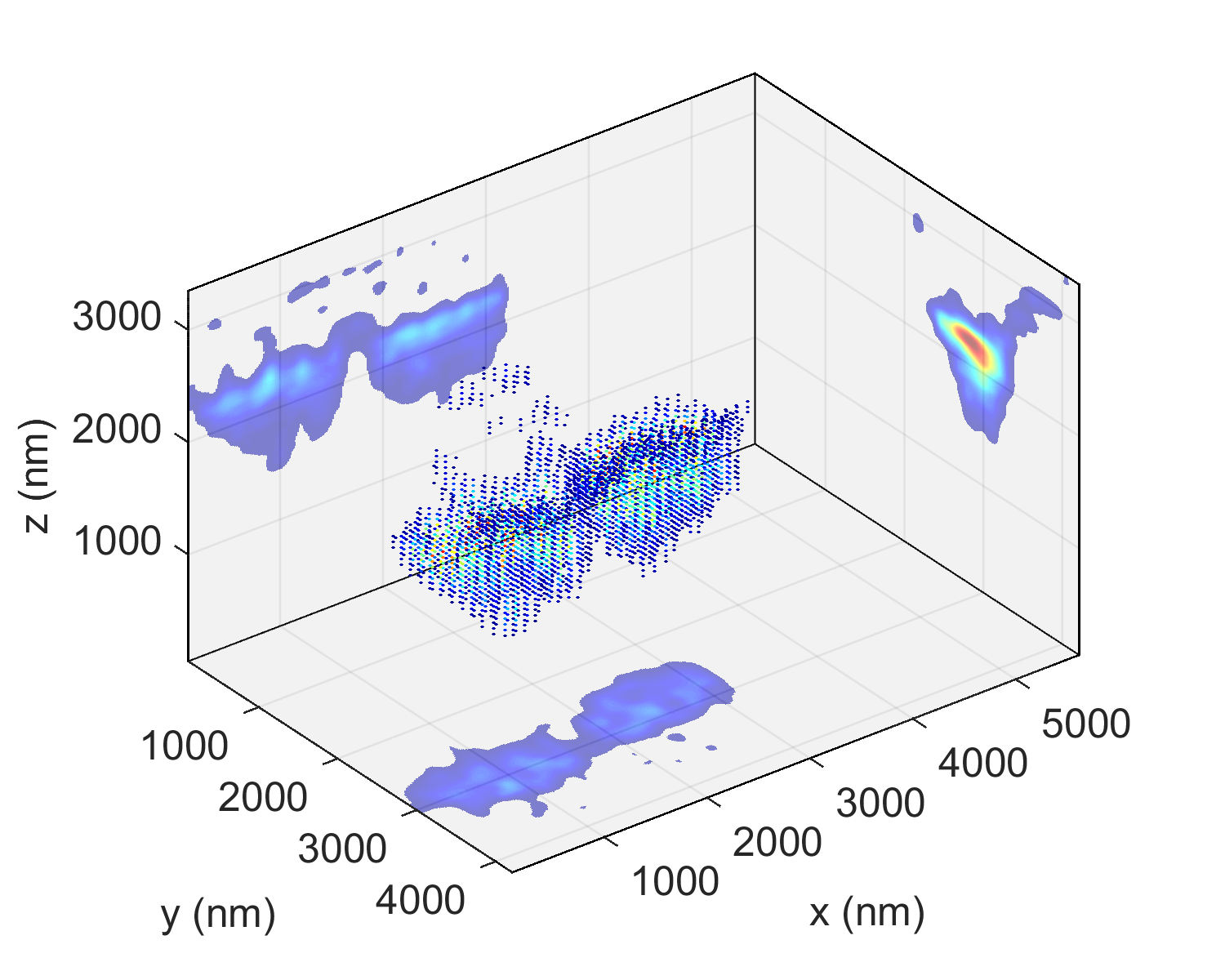}
 \hspace{-0.2in}
 \includegraphics[width=0.11\linewidth]{Figures/colorbar_jet.png}
  \subcaption{3D snapshot MFM} 
\end{subfigure}
\end{minipage}
\begin{subfigure}[b]{0.64\linewidth}
  \centering
  \includegraphics[width=1\linewidth]{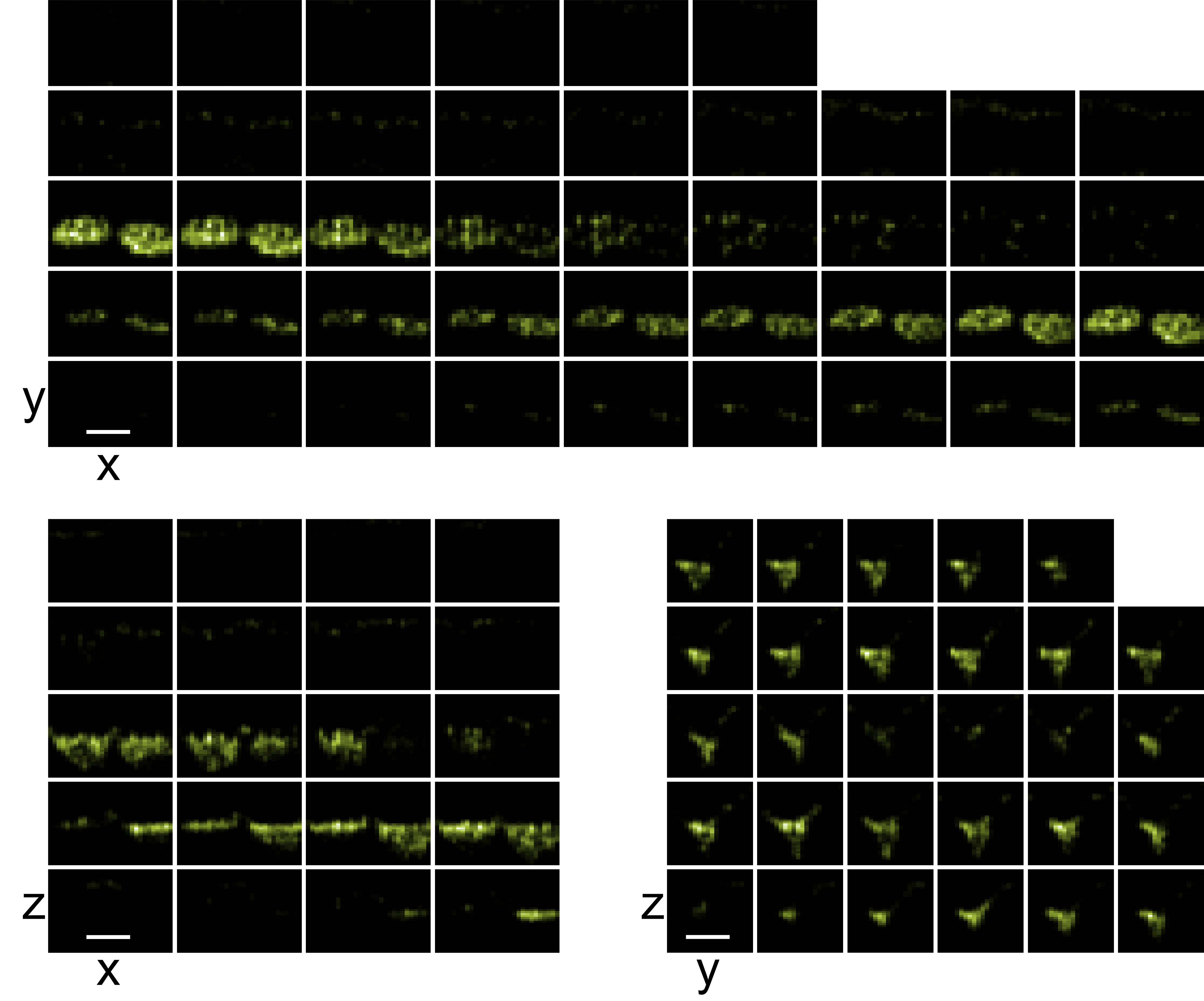}
  \subcaption{Reconstruction sliced along z (top), y (left), and x (right)}
\end{subfigure}
\caption{Results from 3D snapshot MFM imaging. (a)~Point cloud visualization of a bacterium from an experimentally measured stack of confocal images. (b)~Reconstruction of the same bacterium from a 3D snapshot image. 
(c)~Volumetric data from reconstruction presented as sequential slices along each coordinate direction, cropped from a much larger but otherwise empty scene. The scale bar is 1$\mu$m.}
\label{fig:Results}  % Data captured on 20160809
\end{figure*}

Three dimensional reconstructions of fluorescently labeled \textit{Pseudomonas fluorescens} bacteria using this method compare well with volumetric data from a confocal microscope of the identical bacteria deconvolved with its point spread function (PSF) (figure~\ref{fig:Results}(a) and (b)).   
The 3D snapshot data acquisition was over 600 times faster than the confocal data acquisition time; in this instance, 40~ms compared to $\sim$26~s. 

\begin{figure*}[ht!]
\begin{minipage}[b]{0.44\linewidth}
\centering
\begin{subfigure}[b]{1\linewidth}
  \includegraphics[width=0.87\linewidth]{Figures/20160809_Bacteria1_DoubleLobes_fCF.png}
  \hspace{-0.2in}
  \includegraphics[width=0.11\linewidth]{Figures/colorbar_jet.png}
  \subcaption{Confocal $z$-stack}
\end{subfigure}

\begin{subfigure}[b]{1\linewidth}
  \includegraphics[width=0.87\linewidth]{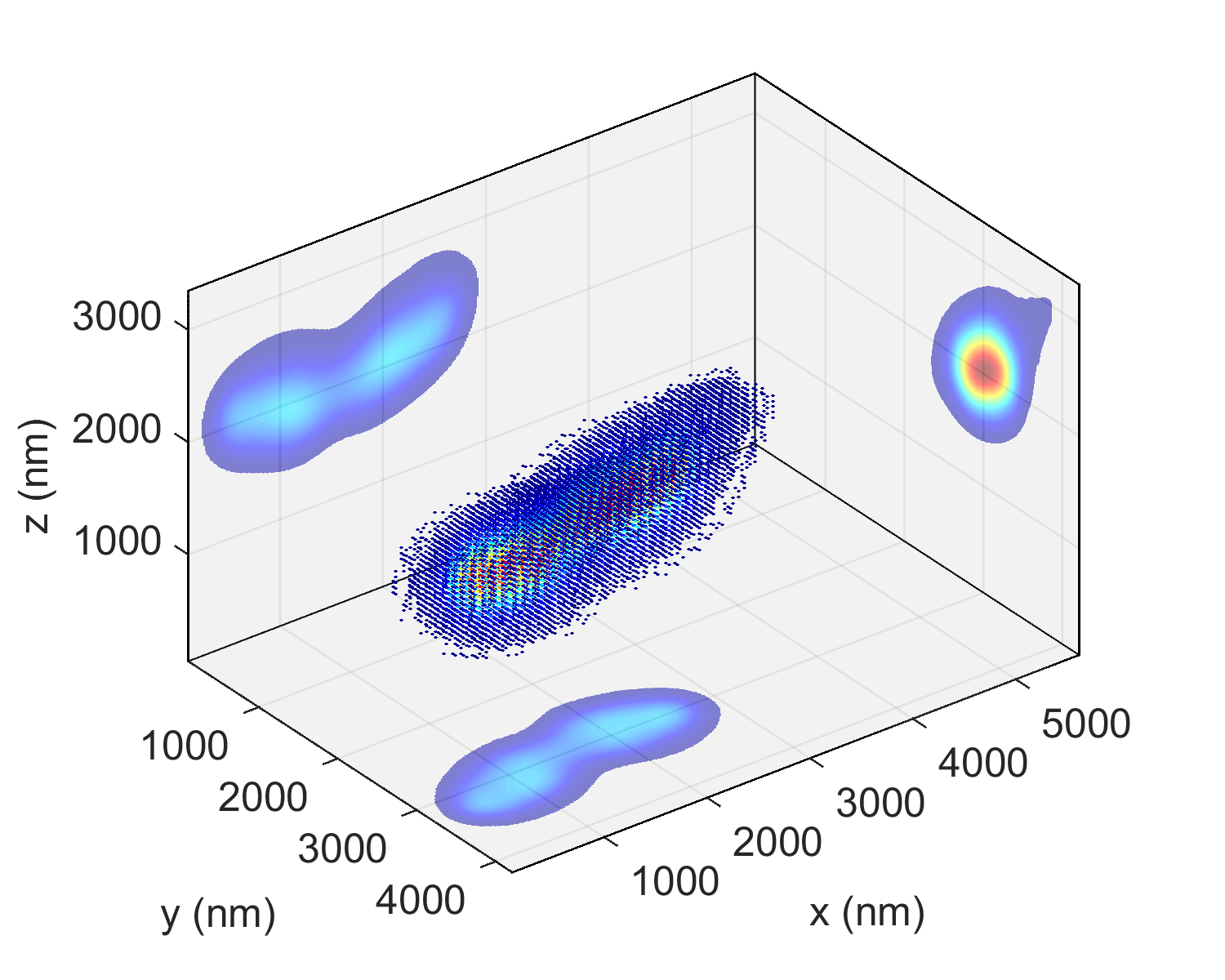}
  \hspace{-0.2in}
  \includegraphics[width=0.11\linewidth]{Figures/colorbar_jet.png}
  \subcaption{Reconstructed volume }
\end{subfigure}

\begin{subfigure}[b]{1\linewidth}
  \centering
  \includegraphics[width=1\linewidth]{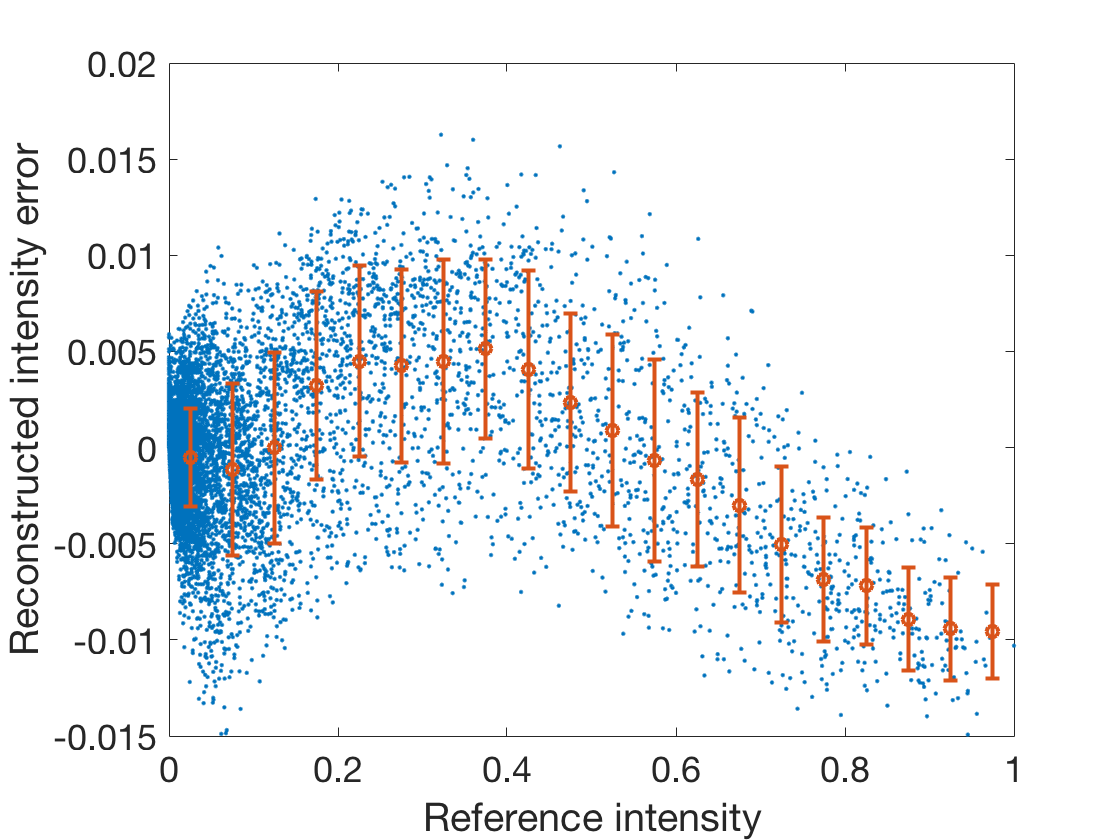}
  \subcaption{Voxel-by-voxel comparison}
\end{subfigure}
\end{minipage}
\begin{minipage}[b]{0.55\linewidth}
\centering
\begin{subfigure}[b]{1\linewidth} % these can be as big as 0.49\linewidth
  \centering
  \includegraphics[width=1\linewidth]{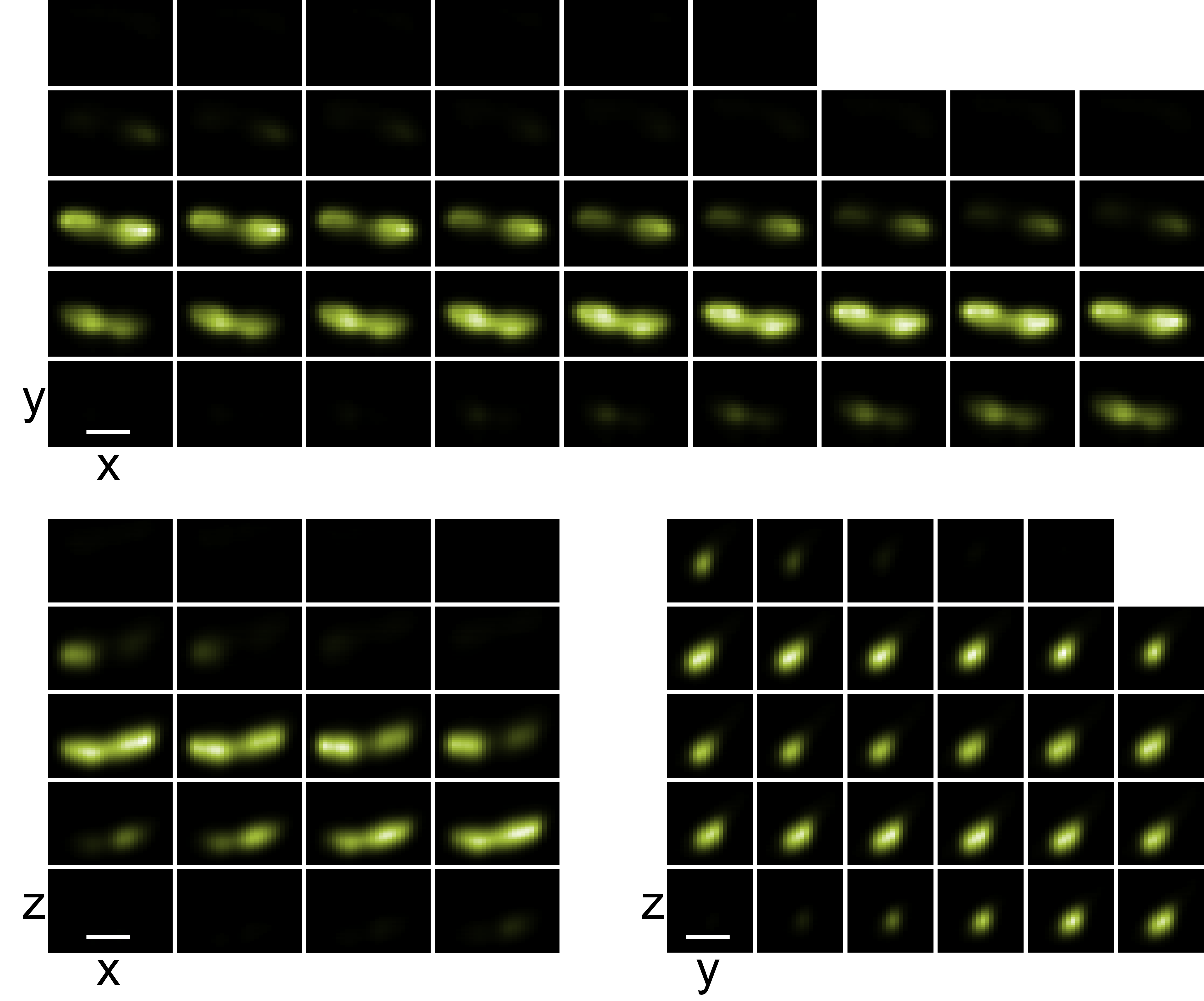}
  \subcaption{$x$-, $y$-, $z$-slices of reconstructed volume}
\end{subfigure}

\vspace{0.2in}
\begin{subfigure}[b]{1\linewidth}
  \centering
  \includegraphics[width=1\linewidth]{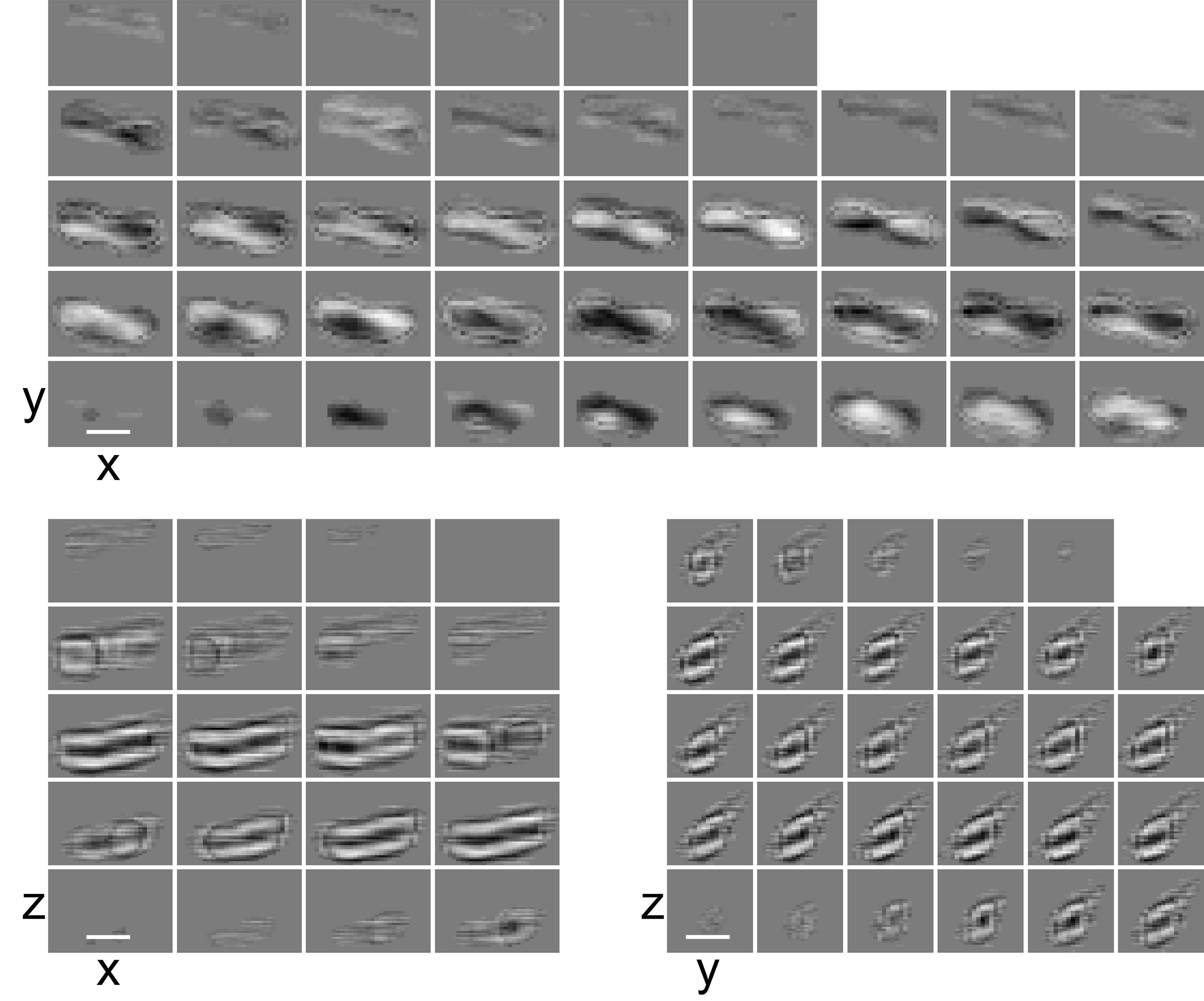}
  \subcaption{Reconstructed intensity error ($\pm$ 0.015)}
\end{subfigure}
\end{minipage}
\caption{3D Reconstruction validation results.
(a)~Reference volume measured using confocal microscope.
(b)~Reconstruction from simulated MFM snapshot derived from the reference volume, resulting in a PSNR of 47.33 db.
(c)~Voxel-by-voxel comparison of reconstructed volume with the reference volume shown as the difference vs. the reference intensity (blue dots). Red indicates mean and standard deviation of residual values in each reference intensity bin.
(d)~Slice-by-slice presentation of the volume reconstructed from simulated MFM images of bacterium.
(e)~Residual between reconstructed and reference volumes.
} \label{fig:RVreconstruction} % Data captured on 20160809, 20161012_1626_simulated_sigmaG=4.36e-04_sigmaPSF=6.28e-04_nVoxel=9091__wReg=1.0e+00
\end{figure*}

To validate our method, we first created a reference volume using confocal imaging of a single bacterium. This data was resampled to 108 nm $\times$ 108 nm $\times$ 50 nm. We simulated an MFM snapshot of the bacterium, as seen in figure~\ref{fig:RVreconstruction}, by applying the forward model to this resampled volume. The 3D reconstruction from this simulated snapshot provides us with a measure of the ideal performance of the algorithm in the absence of added noise.  
Analysis and visualization (figure~\ref{fig:RVreconstruction}(c-d)) of the voxel-by-voxel comparison of the reference volume to the 3D reconstruction demonstrate that each voxel value can be interpreted as a measurement of the intensity in the sample accurate to 1\% of the peak intensity.

\begin{figure*}[ht!]
\centering
\begin{subfigure}[b]{.50\linewidth}
  \centering
  \includegraphics[width=1\linewidth]{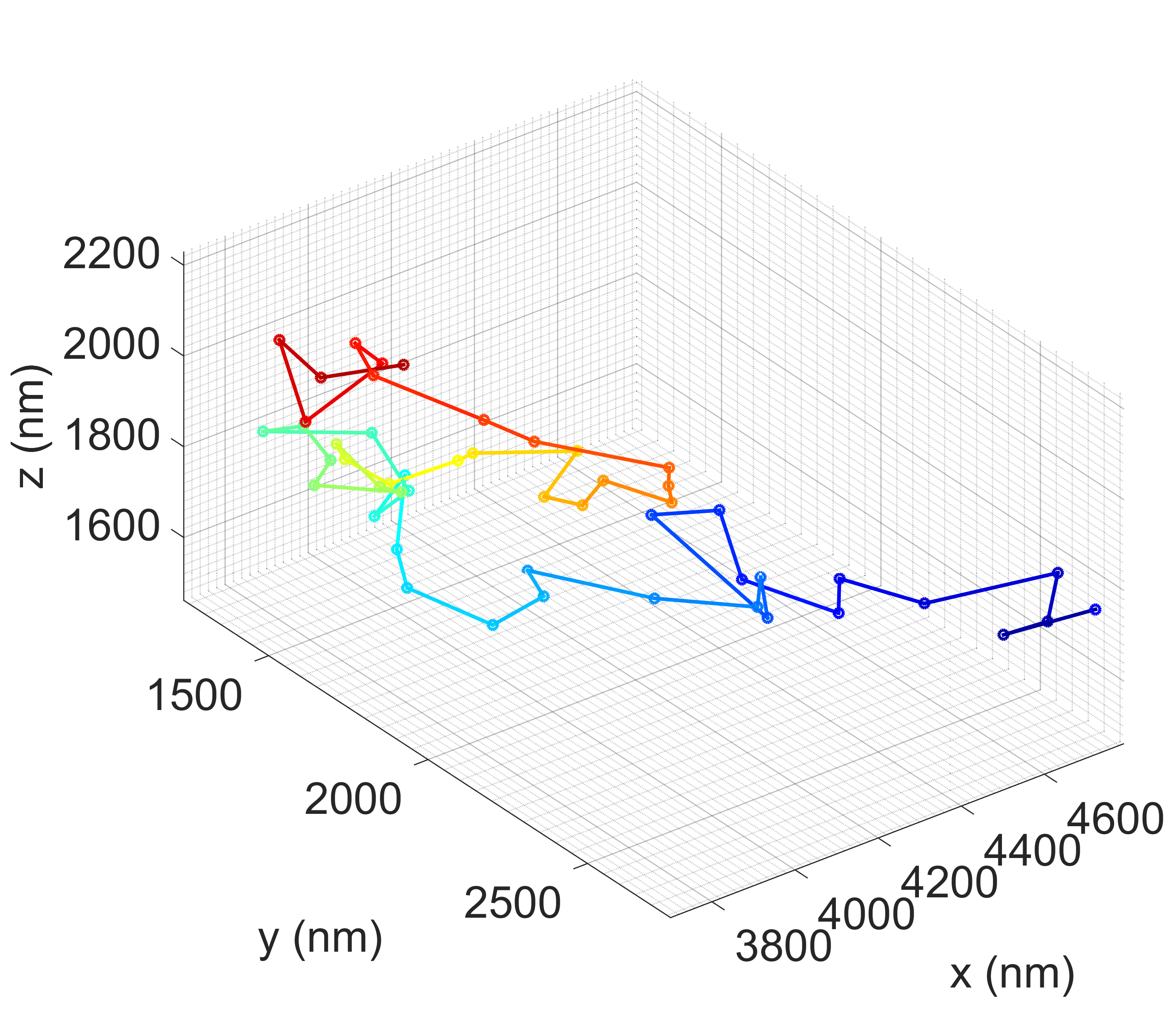}
   \subcaption{Trajectory of centroid}
\end{subfigure}
\hspace{0.1in}
\begin{subfigure}[b]{.47\linewidth}
  \centering
  \includegraphics[width=1\linewidth]{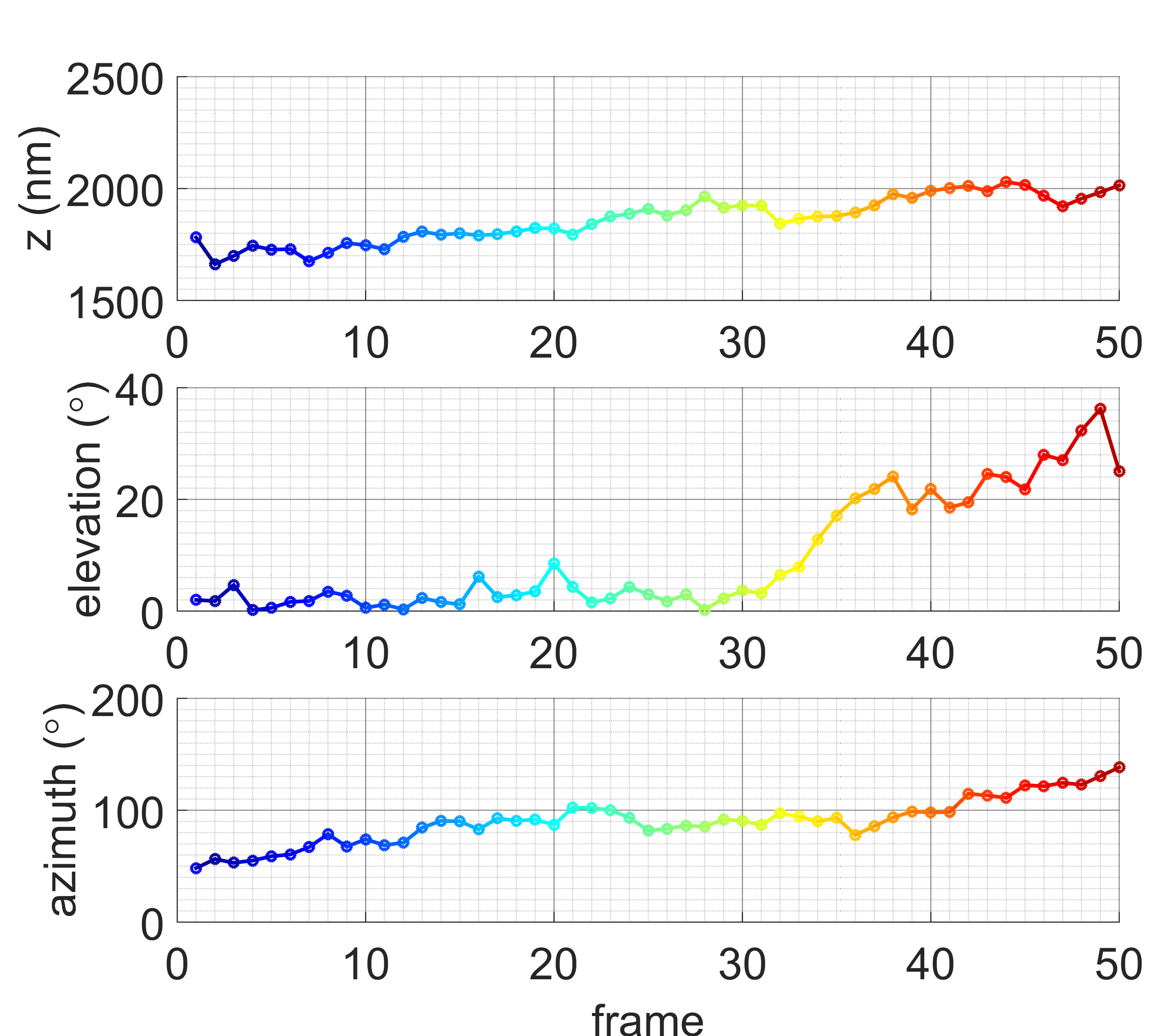}
    \subcaption{z-position, elevation, azimuth}
\end{subfigure}
\caption{Tracking a bacterium following image reconstruction of MFM data diffusing in a coverslip sandwich sample cell for 50 frames acquired at 25 frames per second: (a) trajectory of the centroid of the reconstructed bacterium, and (b) $z$-position of centroid, elevation and azimuth angles.}
\label{fig:Tracking}  % Data captured on 20160809
\end{figure*}

The imaging rate, 
signal to noise ratio (SNR),
and robust reconstruction of the method allowed tracking the 3D motion of a chemically fixed bacterium (i.e., not live) diffusing in water to demonstrate its ability to capture dynamics of rigid body motion of a biological sample (movie S1). 
Figure~\ref{fig:Tracking} (a) plots the 3D coordinates of the centroid of the bacterium over 50 frames (2 seconds) and (b) plots the $z$-position and the elevation and azimuth angles of the major axis of the bacterium. Its major axis remains parallel to the $x$-$y$ plane (elevation angle near zero) for the first 30 frames while the bacterium turns freely about the $z$-axis (azimuth angle).  As the centroid moves upward in $z$, presumably away from the coverslip, the bacterium is able to tilt upward, increasing the elevation angle of its major axis.
We believe this is the first demonstration of 3D volume reconstruction at these frame-rates for extended objects.

\section*{Discussion}

The 3D snapshot MFM imaging system simultaneously captures a full 3D dynamic fluorescent scene of approximately 20 $\mu$m $\times$ 20 $\mu$m $\times$ 10 $\mu$m in size at 25 Hz rendered with 108 nm $\times$ 108 nm $\times$ 50 nm voxels.
Its snapshot capability distinguishes it from conventional microscopies, which is of particular significance for studying high speed dynamical processes.
Computational imaging in this system enables a relatively inexpensive alternative to complex optical elements while extending the reach of MFM to include capture of complex scenes including extended objects and confusion-limited aggregates of point sources.
Further development of this method to improve resolution and accuracy by better characterization of the instrumental PSF, improved accounting of the image noise model, and improved modeling of the image priors will further enhance its potential to contribute to the study of dynamical processes at very high spatio-temporal resolution.

\section*{Methods}

The following sections give details on the experimental and mathematical methods in support of the results presented in the paper.

\subparagraph*{Multifocal Microscope.}
Our multifocal microscope was built around a Nikon Eclipse Ti inverted microscope and is outlined in Figure~1(a) in the main text. 
Our instrument is based on the design in~\cite{Abrahamsson2012}.  
In a significant departure from that design, we discard the aberration correcting elements--the chromatic correction grating and custom multifaceted prism and address aberration correction computationally in image reconstruction.
The sample (S) was mounted on a mechanical stage (Ludl 99S106-N2K- LE2, controller MAC5000). 
Epifluorescence emission collected through the objective (OBJ, Nikon 60x 1.27 NA CFI Plan Apo water immersion, MRD07650) was separated from excitation light by a dichroic mirror (DM, Semrock 488 LP, Di03-R488-t3-25x36) and bandpass filter (BP1, Semrock 520/28, FF02-520/28-25).
Excitation light was bandpass filtered (EX, Semrock 472/30, FF02-472/30-25) from the cyan channel (196 mW LED) of a Spectra X light engine (Lumencor).  
A field stop (FS) crops the field of view to 1~mm$^2$. 
The field stop consists of a circular $25.4$~mm diameter by 1~mm thick fused silica substrate (Thorlabs WG41010) with a 50~nm Cr layer deposited on the entire surface except for a 1~mm square in the center. 
A 4-f lens system relays the light from the side port onto the camera (Andor iXon Ultra 888 EMCCD) with an additional 2x magnification for a total system magnification of 120x. 
The first lens (RL1, f = 200~mm, Thorlabs AC508-200-A) matches the focal length of the tube lens (TL, Nikon, f = 200~mm) and images the secondary pupil plane of the objective onto the custom-made (see below) MFM diffractive optical element (DOE). 
The second lens (RL2, f = 400~mm, Thorlabs AC508-400-A) images the diffracted and shifted fields of view from the DOE onto the camera (EMCCD). 
A second, narrower bandpass filter (BP2, Semrock 520/15, FF01-520/15-25) is placed immediately before the camera to narrow the bandwidth of the emission light 
to mitigate chromatic aberration.
The 120x magnification gives a theoretical lateral pixel size of 108~nm, which was confirmed by imaging a ruled slide.
The lateral field of view in each MFM tile is 16.6 ${\mu}$m by 16.6 ${\mu}$m
(Figure~1a, main text).

\subparagraph*{MFM Diffractive Optical Element Design and Fabrication.}
The MFM DOE was designed with an undistorted pitch of 91 ${\mu}$m and a focal shift ${\delta}z = 250$~nm between tiles following a procedure outlined by Abrahamsson et al.\cite{Abrahamsson2012}. The result of the iterative algorithm is a binary grating function representing phase shifts of 0 and $\pi$ at each pixel.

\begin{figure}[ht!]
\begin{tabular}{cc}
\includegraphics[width=0.45\linewidth]{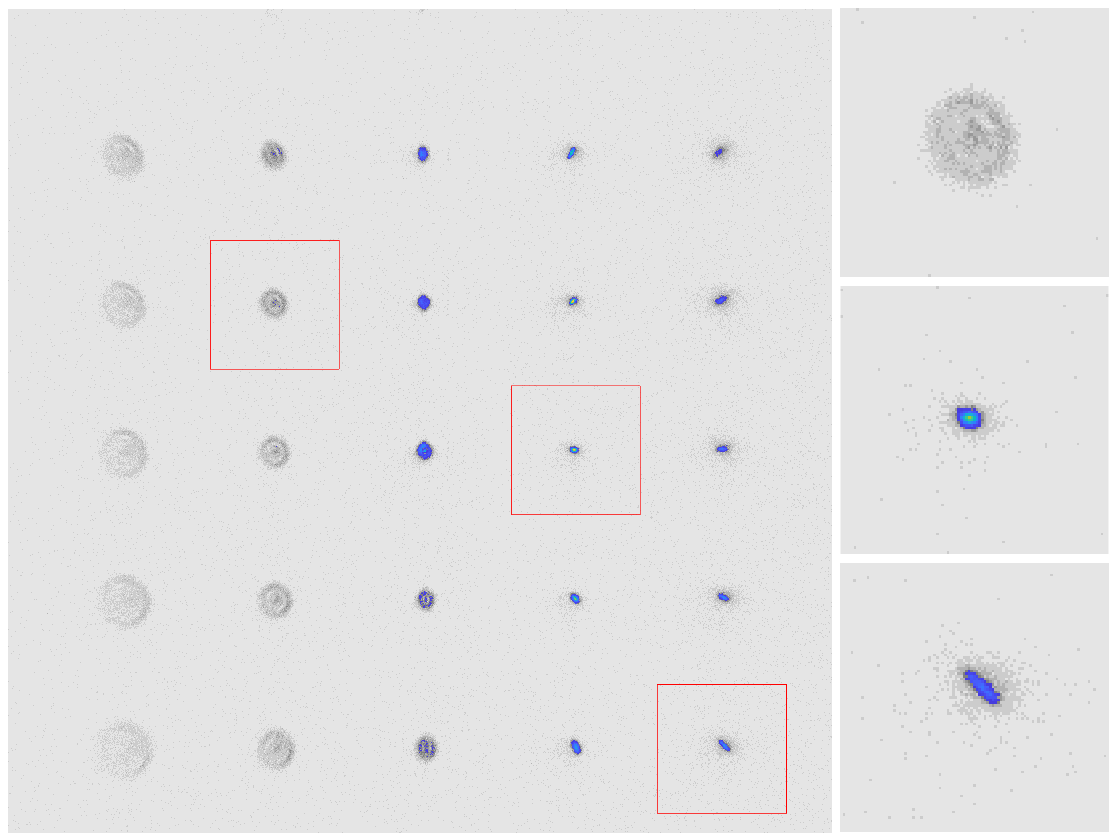}&
\includegraphics[width=0.45\linewidth]{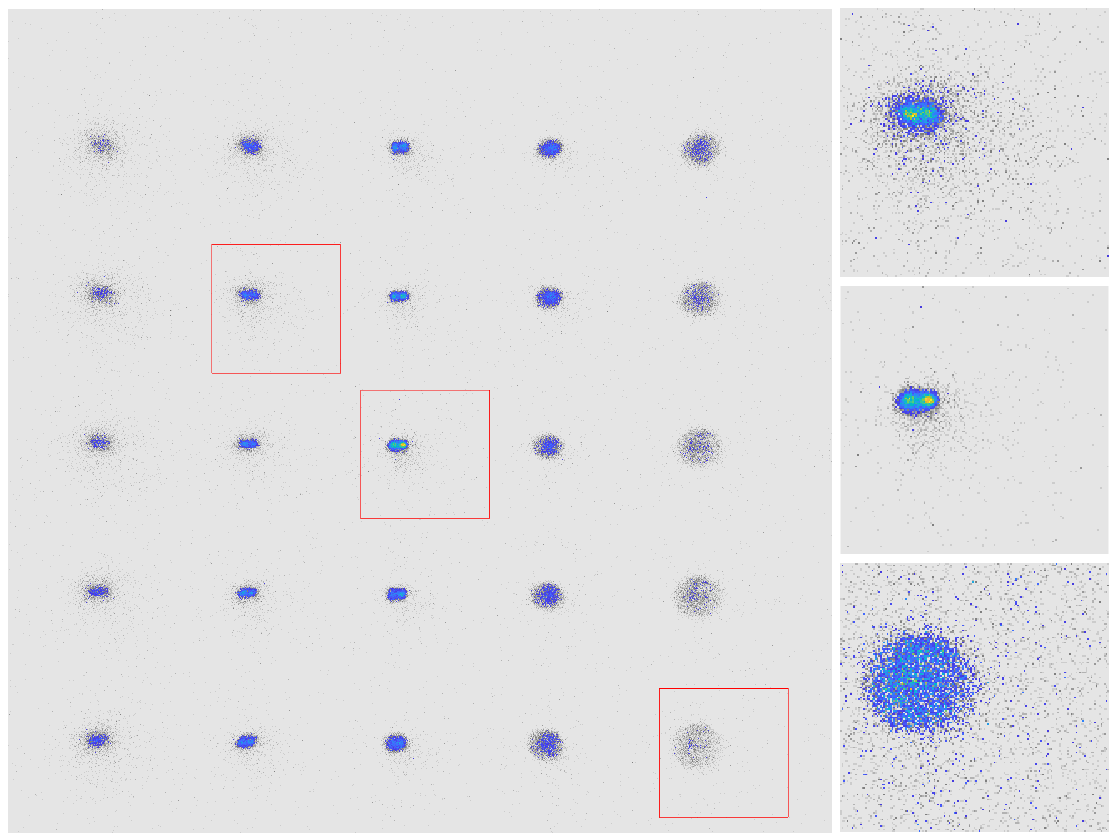}\\
(a) & (b)\\
\end{tabular}
\caption{Example of MFM images obtained from  (a) a 170~nm fluorescent bead and (b) a fluorescently  labelled bacterium.
The lowest intensity bin of the false color map has been set to light gray for visibility.
The side images contain detail of the corresponding indicated regions.
Notice that the bead is nominally in focus in 3rd row of the 4th column of the MFM image array.
The bacterium is approximately in focus in the center tile.
The focal shift between tiles is set by the DOE design at 250~nm,
so that the 25 tiles in  MFM imaging span a focal shift from -3 $\mu$m in the upper left corner
to +3 $\mu$m in the lower right corner.
} \label{fig:MFMexample}
\end{figure}

Our grating function produces a 5 by 5 grid and has a diffraction efficiency of 78\%.
In order to construct the DOE, a 5-mm thick UV fused silica substrate (Thorlabs WG41050) was cleaned by using an oxygen plasma and spin-coated with a 2 ${\mu}$m layer of S1813 photoresist (Shipley). 
A laser writer (Heidelberg) exposed
the MFM grating pattern into the photoresist (402~nm laser at a dose of 80 mJ). 
The photoresist was developed in AZ-300 MIF developer (Integrated Micro Materials) for 20 s. 
Then the pattern was etched in a fluorine reactive ion etcher (RIE, Oxford Instruments). 
The etch depth of 578~nm for the design wavelength of 525~nm was etched by using gases Ar (25 standard cubic centimeters per minute (sccm)) and $\textrm{CHF}_3$ (25 sccm), an RF power of 200 W, and inductive plasma coupling power of 250 W (etch rate of 32~nm/min, 1:7 photoresist to fused silica selectivity). 
The  photoresist remaining after etching was stripped with acetone in an ultrasonic bath. 
The mean amplitude of the surface roughness of the finished diffractive optic was determined to be $\sim\lambda/8$ using contact profilometry. 
All fabrication steps were completed at the Pritzker Nanofabrication Facility at the University of Chicago. 
The MFM grating produced with this procedure is shown (inset) in Figure~1a.  
Examples of the resulting MFM image structure (Figure~\ref{fig:MFMexample}) show the placement of 25 focal shifted sub-images folded into a 5 $\times$ 5 grid of tiles on the detector.

\subparagraph*{\textit{Pseudomonas fluorescence} SBW25 Strain Modification and Sample Preparation.}

SBW25, isolated from sugar beet leaf in 1989, is widely studied as a model system of plant colonization and plant-microbe interactions~\cite{bacteriaprep}. 
Its genome has been sequenced and analyzed by Silby et al.~\cite{Silby2009}.
SBW25 has plant growth promoting properties and acts as a biocontrol agent, mediating systemic resistance to bacterial, fungal, and oomycete pathogens (for example, Trippe et al.~\cite{Trippe2013}; Naseby et al.~\cite{Naseby2001}). 
We aim to study the dynamics, organization, and interactions of SBW25 with plant roots and other microbial species in the rhizosphere.
To achieve this objective, we have labeled SBW25 with a rhizosphere-stable plasmid expressing the fluorescent protein mNeonGreen~\cite{Shaner2013}. 
The plasmid is a derivative of pME6031~\cite{Heeb2000} and was adapted for our use by introduction of a constitutive promoter for strong expression of the fluorescent protein. 
The plasmid is stably maintained in SBW25 during rhizosphere colonization 
permitting bright labeling of SBW25 in the absence of selective pressure.

SBW25 bacteria expressing unconjugated mNeonGreen were prepared fresh by transferring a colony from an agar plate with a 1 ${\mu}$L sterile loop to 5 mL of nutrient broth. 
The bacteria then were grown in the dark for 10 hours at room temperature with gentle agitation. 
After growth, the bacteria were chemically fixed as previously described~\cite{bacteriaprep2,Huynh2017} to preserve the mNeonGreen fluorescence.

\begin{figure}[ht]
\centering
\includegraphics[width=0.95\linewidth]{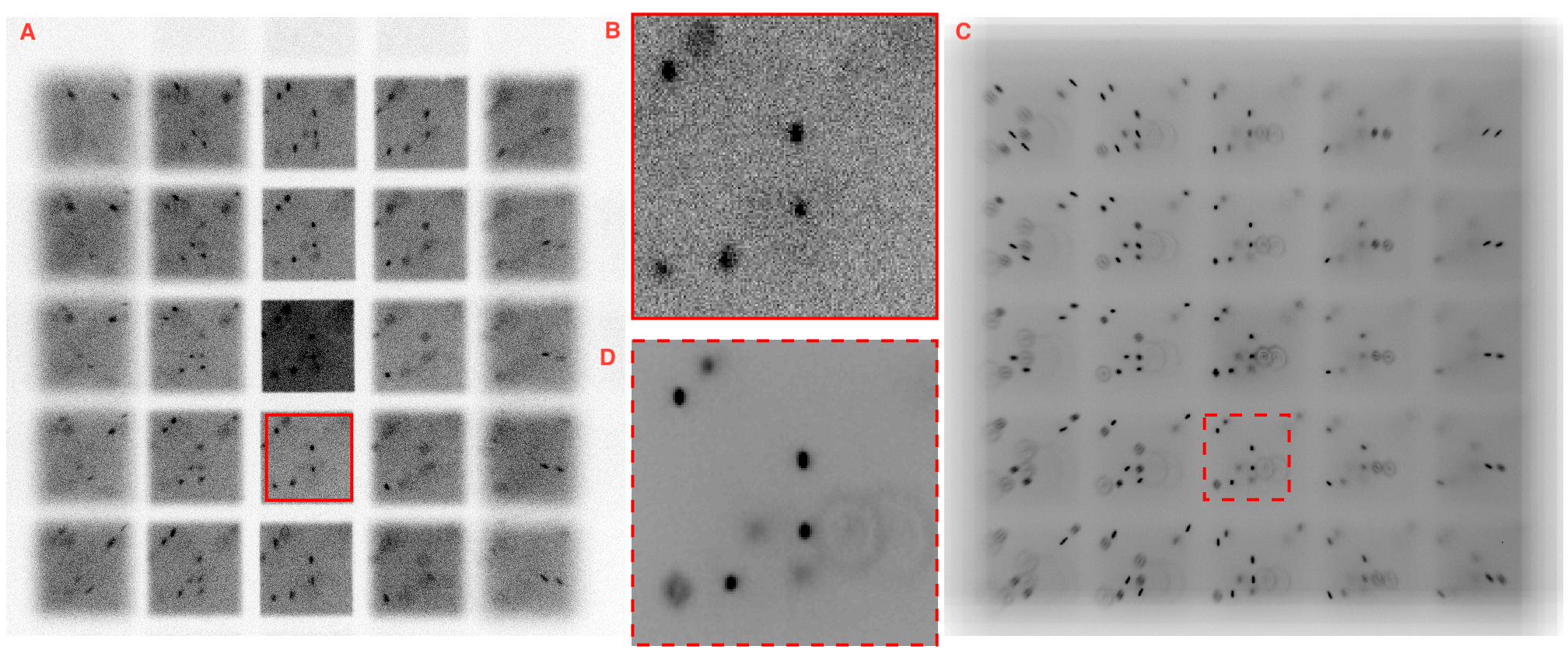}
\caption{
MFM imaging of fluorescent beads. (a) The MFM image of a sample containing multiple fluorescent beads at different locations and depths; (b) a detail of one tile .
The image in (c) is generated by our forward model for a volume with 15 point sources.
(d) A corresponding detail of one tile.
We used the image on the right to guide our choice of bead location.  Notice that some of the observed light in (b) came from beads outside the volumetric field of view.
}
\label{fig:ForwardModelExample}
\end{figure}

All samples were prepared on a $22 \times 22$~mm No. 1.5H ($170\pm 5$ ${\mu}$m thickness) glass coverslip (Bioscience Tools, San Diego, CA), mounted onto a 1~mm thick glass slide (Fisher Scientific), and sealed with nail polish (Electron Microscopy Sciences, Hatfield, PA).   
For PSF measurements (Figures~\ref{fig:MFMexample}a and \ref{fig:ForwardModelExample}), 170~nm carboxylated polystyrene beads infused with green fluorophores (Invitrogen) were diluted 100-fold in poly-l-lysine (Sigma). 
A 1 ${\mu}$L drop was pipetted on the coverslip. The coverslip was mounted as described above. 
The coverslips were plasma cleaned for samples of stationary bacteria (Figures~2 and 3 in the main text, Figure~\ref{fig:MFMexample}b)%, coverslips were plasma cleaned.  
After cleaning, three fiducial marks were made on the sample side of the coverslip with permanent marker in order to be able to relocate the same bacterium between the MFM and confocal microscopes~\cite{Huynh2017}.
Then, a 1 ${\mu}$L drop of the diluted (M9 buffer) fixed bacteria suspension ($\sim$20,000 bacteria) was pipetted onto a coverslip, and the coverslip was mounted as described above.  
For diffusing bacteria samples (Figure~4, main text), a 10 ${\mu}$L drop was pressed between the coverslip and glass slide. 
The sample was prepared immediately before imaging, because the untreated glass would slowly adsorb the bacteria.

\subparagraph*{Confocal Microscopy and Correlative MFM-Confocal Imaging. }
Confocal microscopy images were acquired with a 100x 1.45 NA oil immersion objective (Nikon 100x NA 1.45
Plan Apo oil immersion, MRD01905) through a Yokogawa W1 spinning disk (CSU-W1, 50 ${\mu}$m pinhole;
Solamere Technology Group, Salt Lake City, Utah) attached to a Nikon Eclipse Ti. The W1 was equipped with
a custom 1.83x magnification lens insert (Solamere Technology Group) for a total magnification of 183x. The
resulting pixel size was 71~nm on the detector (Andor iXon Ultra 888 EMCCD) as measured with a ruled slide.
mNeonGreen fluorescence excitation employed a 488~nm laser diode (Spectra-Physics Excelsior One 488C-100) and
emission was filtered by a dichroic beamsplitter (Semrock, Di01-T405/488/568/647-13x15x0.5) and emission
filter (Chroma, ET525/50m). Data collection and system control were managed by ${\mu}$Manager
software\cite{Edelstein2014}.

Correlative imaging between MFM and confocal microscopy was accomplished by using the FARMER
method~\cite{Huynh2017} to relocate the field of view from the MFM mechanical stage to the Prior
mechanical stage (Prior stage H117E1N5/F, controller H31XYZ7EF) on the confocal microscope. Several
bacteria with unique shapes and in sparse locations were imaged by using MFM, and their \textit{x, y, z} coordinates were
recorded with ${\mu}$Manager.
The fiducial marks on the cover slip were imaged  using brightfield
microscopy and the \textit{x, y, z} coordinates were recorded. The coordinates provided by FARMER were used to relocate each
bacterium. The unique shapes and sparsity of the sample allowed for correction of small errors in the
relocation.  The confocal images stacks were acquired with a $z$ step of 50~nm, an EM gain of 250, and an exposure time of 200 ms.
These image stacks were used as part of validation of the computational image reconstruction method we developed.

\subparagraph*{Imaging Acquisition Parameters.}

For diffusing bacteria (Figure~4, main text), the detector was operated at the hardware maximum 25 Hz frame rate (full frame) with an EM gain of 760.  
The motion of the bacteria induced motion blur when using a 40 ms exposure, as in previous experiments.
In order to further reduce motion blur, the exposure time was reduced by strobing the excitation light synchronously with camera exposure with a $50\%$ duty cycle (i.e., illuminated for 20 ms of a 40 ms exposure) using custom software written in Labview.

\subparagraph*{MFM Forward Model.} \label{sec:ForwardModel} %Mathematical Model of MFM

We present a mathematical model describing the formation of the 2D MFM image
from the 3D object volume. This forward model incorporates the physical elements of
the system and provides a tool to be used both in our reconstruction algorithm, to
recover a 3D object volume from a measured 2D MFM image, and in our simulation
experiments, to generate a simulated 2D MFM image from a given 3D object volume.

The MFM forward model links a 3D volumetric fluorescence distribution in object space (e.g., from a bacterium)
with its corresponding 2D MFM image in the microscope's detector space.  We discretize the
sample volume into $N_x \times N_y \times N_z$ voxels and denote by $f(x,y,z)$ the intensity of voxel $(x,y,z)$.
Because the emitted fluorescent light is incoherent, the image formulation
process is linear in power; hence, for each tile $t=1, 2, \cdots,25$,
the observed 2D image $g_t\left(u, v\right)$ of the volumetric object is a linear superposition.

The imaging system is shift invariant within each tile so that the image formation model
can be expressed as the $x,y$ 2D convolution of the
source $f(x,y,z)$ volume with the PSF tile $\psf_t(u,v; z)$ and then summed along $z$,
\begin{equation}
g_t\left(u, v\right)
=\sum\limits_{z=1}^{N_z} \sum\limits_{x=1}^{N_x} \sum\limits_{y=1}^{N_y} {\psf_t(u - x, v - y; z)f(x,y,z)}, \label{eq:tileForwardLinearShiftInv}
\end{equation}
where $t=1, 2, \cdots,25$.
We have verified the intra-tile spatial shift invariance both by simulation and experiment. This property is essential in order to quickly measure the PSFs and develop efficient reconstruction algorithms that utilize fast Fourier transforms (FFTs).
Figure~1b (main text) is a pictoral representation of Equation~(\ref{eq:tileForwardLinearShiftInv}). % pictorially.

In our computations we set the $z$ dimension of the voxel as 50 nm to match the step
size used when measuring the PSF, and we set the $x$ and $y$ dimensions (108 nm) %(both 108 nm)
of the voxel to match the
sensor's pixel pitch ($13$ $\mu$m)
divided by the magnification $M=132$.
Those choices are convenient because they remove the need for resampling
when applying the forward model.
The most appropriate choice of resolution parameters (voxel dimensions) will generally depend on the signal-to-noise ratio,
number of focal planes sampled by the MFM snapshot
(25 in our setup), and step size used in measuring the PSF.

To simplify the notation and formulate the volume reconstruction as a optimization problem, we represent
$\bm{f}\in \bm{R}^{N_xN_yN_z}$ and $\bm{g} \in \bm{R}^ {M_xM_y}$ in lexicographic vector form~\cite{Katsaggelos2007}
from the 3D object voxels $f(x,y,z)$ and the entire set of 2D sensed
pixels $g_t(u,v)$ for each tile.\footnote{A 2D image or 3D volume can be laid out as a single vector.  In MATLAB\texttrademark syntax, for example, the volume f(:,:,:) can be represented lexicographically by using the expression f(:).}
We can now rewrite Equation~(\ref{eq:tileForwardLinearShiftInv}) as the linear matrix vector multiplication,
\begin{equation}
\bm{g} = \bm{Af}, \label{eq:linearForwardMatrix}
\end{equation}
where the elements of the coefficient matrix $\bm{A}\in \bm{R}^{M_xM_y \times N_xN_yN_z}$ are a
reorganization of the elements of the PSF for each tile $\psf_t(u, v; z)$. Note that the matrix $\bm{A}$ characterizes the effects of the
elements of the optical system including the DOE, wavelength of light, pixelization at the detector,
field stop, and lenses. Errors in the forward model arise from multiple sources such as noise in
the measured PSF, errors in the PSF, and errors in modeling the effects of the field stop.

Figure~\ref{fig:ForwardModelExample} demonstrates the ability of the forward model to represent a measured MFM image of a 3D scene.
A volume containing 15 point sources is used as input to our forward model. The image on the left side shows the forward model for this volume.  The point source locations are selected based on the MFM image shown on the right.

\subparagraph*{Computational Reconstruction of a 3D Object Image from a 2D MFM Image.}
\label{sec:Reconstruction}

Our goal is to reconstruct 3D information about biological objects from MFM images.
Previous MFM systems~\cite{Ram2008} have developed approaches that localize \emph{point} sources.
These approaches will not be applicable to recovering extended objects, such as unresolved distributions of fluorescent molecules with a bacterium.
Here we present an approach that will recover the detailed distribution of fluorescence in an extended object.
Using the forward imaging model, we restate our goal using the
terms in Equation~(\ref{eq:linearForwardMatrix}).
We treat our MFM measurements as compressive (i.e., $\dim( \bm{f}) \gg \dim(\bm{g})$) so that inversion of Equation~(\ref{eq:linearForwardMatrix}) is an ill-posed problem. As is common for ill-posed problems, we recover the 3D volume $\bm{f}$ by finding the solution to the following minimization problem,

\begin{equation}
\label{eq:PNoiseUnConstrained}
\bm{\hat{f}} = \argmin_{\bm{f}} \quad \frac{1}{2}  \left\| \bm{g} - \bm{Af} \right\|^2  + \lambda {J(\bm{f})},
\end{equation}
where the left hand term tries to enforce Equation~(\ref{eq:linearForwardMatrix})
and the right hand term, $J(\bm{f})$, is the regularization
functional that can be selected to encode desirable properties of $\bm{f}$.  Often the regularization term
rewards smoothness, piecewise smoothness, or sparsity in the signal. For example, $J(\bm{f})$ might bias $f$, or
some transformation of $f$ to another domain, $Df$, to be a sparse signal. In the present work,
we tested both the $L1$ norm and the total variation (TV) as regularizers.
The $L1$ norm regularizer
\begin{equation} \label{eq:L1reg}
J(\bm{f}) = \sum\limits_{x=1}^{N_x} \sum\limits_{y=1}^{N_y} \sum\limits_{z=1}^{N_z} {|f(x,y,z)|}
\end{equation}
approximates the sparsity of the signal.
On the other hand, the total variation

\begin{equation} \label{eq:TVreg}
\begin{split}
J(\bm{f}) = \sum\limits_{x=1}^{N_x-1} \sum\limits_{y=1}^{N_y-1} \sum\limits_{z=1}^{N_z-1} (|f(x+1,y,z)-f(x,y,z)| 
&+ |f(x,y+1,z)-f(x,y,z)| \\
&+ |f(x,y,z+1)-f(x,y,z)|)
\end{split}
\end{equation}
promotes sparsity in the signal gradient.

The Lagrange multiplier, $\lambda$, in Equation~(\ref{eq:PNoiseUnConstrained})
is a weighting factor that trades off the fit to the data against the fit to the prior model
of $\bm{f}$ as encoded in $J(\bm{f})$.
Often $\lambda$ depends on the measurement noise and the conditioning of the forward operator $\bm{A}$. Setting $\lambda=0$ produces a least-squares fit to the measurements. Because the problem is ill-posed, however, a least-squares result will produce an error prone and statistically implausible result.
For large $\lambda$, the solution is heavily biased by the prior model.

Once formulated as a minimization problem, equation~(\ref{eq:PNoiseUnConstrained}) can be computed by using optimization tools. 
A number of optimization/minimization computational methods have been developed to solve problems of the form in Equation~(\ref{eq:PNoiseUnConstrained}). 
Here we use the two-step iterative shrinkage/thresholding (TwIST) algorithm~\cite{Beck2009}.

\begin{figure*}[ht!]
\centering
\begin{subfigure}{.45\linewidth}
  \centering
  \includegraphics[width=1.0\linewidth]{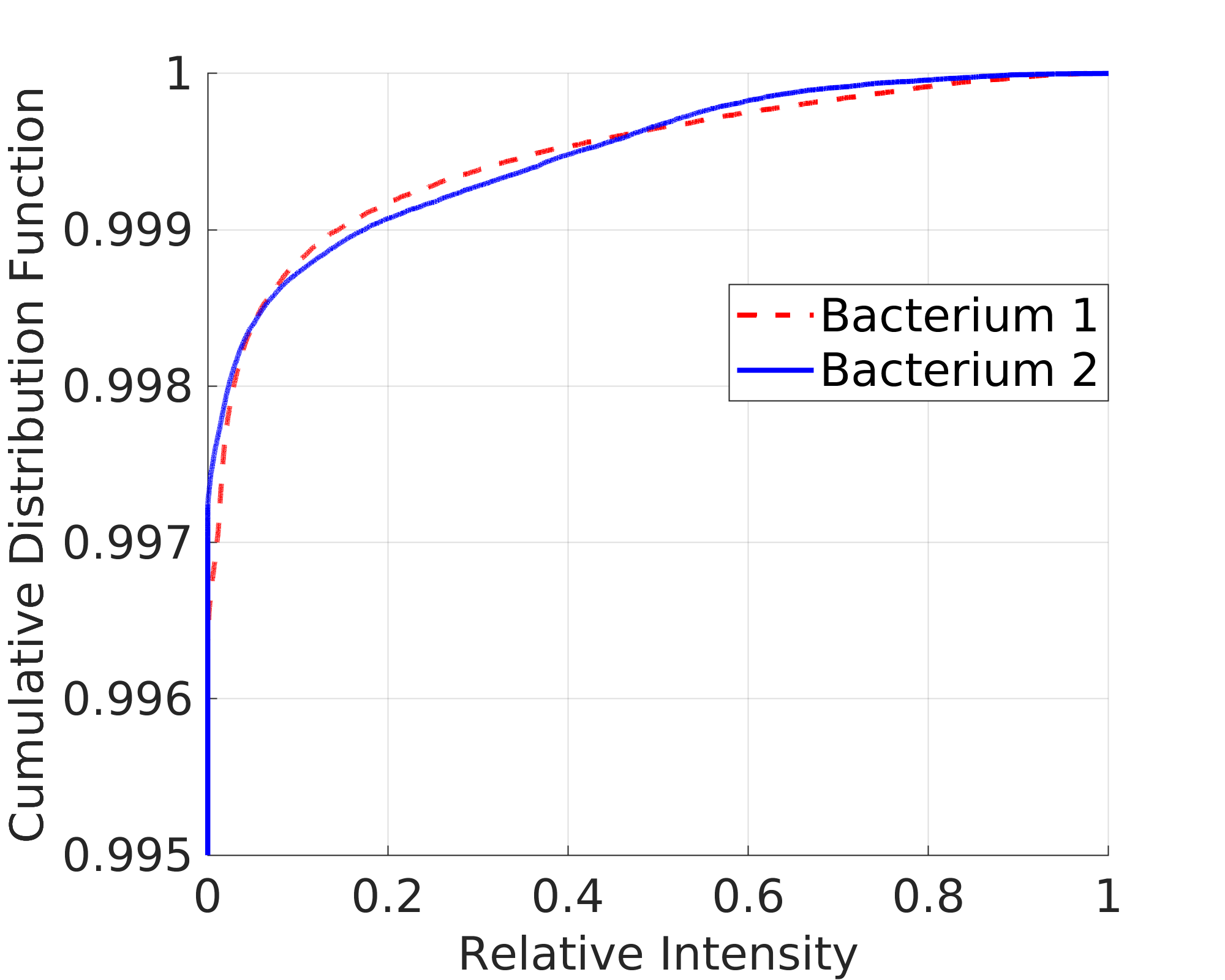}
  \subcaption{Intensity CDF}
\end{subfigure}
\begin{subfigure}{.45\linewidth}
  \centering
  \includegraphics[width=1.0\linewidth]{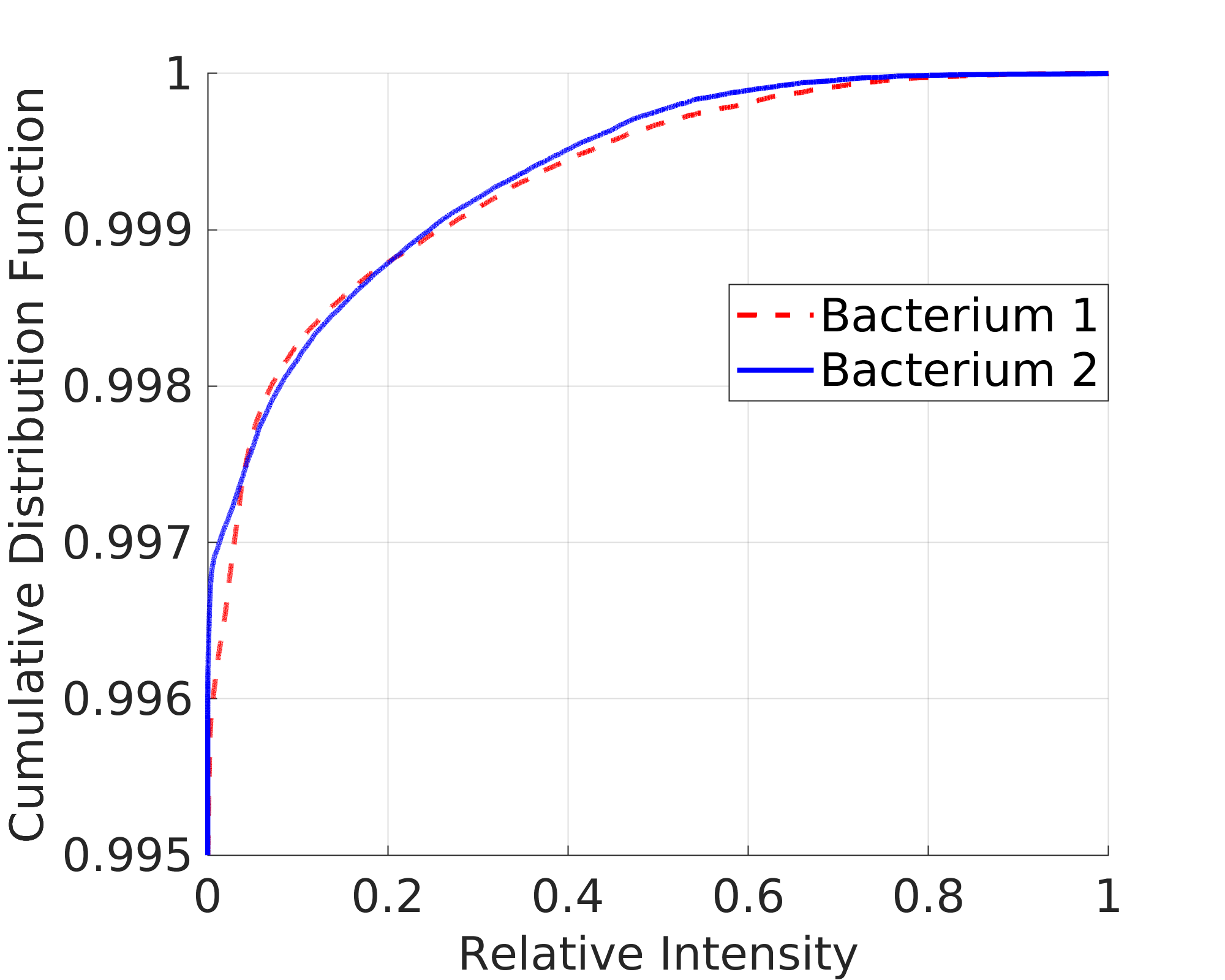}
  \subcaption{Gradient Magnitude CDF}
  \end{subfigure}
\caption{Cumulative distribution function (CDF) of the intensity and gradient magnitude of confocal images of two bacteria (for the entire volume) support the use of a sparsity prior in our regularization approach. (a) shows the CDF of normalized intensity, and (b) shows the CDF of the normalized intensity gradient magnitude.}
\label{fig:cdf} % Data captured on 20160809
\end{figure*}

While many linear systems (e.g. Equation~(\ref{eq:linearForwardMatrix})) can be solved as a minimization problem in the form of Equation~(\ref{eq:PNoiseUnConstrained}), the prior information about the signal as encoded in $J(\bm{f})$ is crucial~\cite{Henderson2013} to the success of the solution as a representation of the real system.
Sparsity is an important assumption because the dimension of $\bm{g}$ is significantly smaller than that of $\bm{f}$.
The question arises, then, whether our observations warrant enforcing this assumption.
Figure~\ref{fig:cdf} presents the cumulative distribution for both the intensity and gradient magnitude of intensity of typical 3D images of two bacteria obtained by using our confocal measurement.
In both cases, all but about 0.3\% of the voxels are essentially zero.
Thus, a sparsity prior using either the L1 regularizer or a TV (gradient) regularizer is suitable for samples of this kind.

\paragraph*{Reconstruction Parameters.} The following summary of the parameters used in reconstruction computations for the results presented in the main text and in Movie S1 to facilitate comparison and to illustrate the variety of circumstances that effect parameter selection.

The confocal image stack presented in Figures 2 and 3 of the main text is comprised of 130 images taken at 50 nm intervals along the axial direction.  Each required 200 ms.  The resulting raw stack was deconvolved with the 3D instrumental point spread function using Huygens software (Scientific Volume Imaging, The Netherlands, http://svi.nl).  The images in this enhanced volume were re-sampled from the native 71 nm pixels of the to match the 108 nm pixel pitch of the MFM measurements.

The MFM snapshot used for the reconstruction in Figure 2 of the main text was captured in 50 ms. The reconstruction used $\lambda=0.01$ for 1000 iterations, reaching the point of diminishing return determined by the level of noise in the image. To compensate for the finite size of the fluorescent bead used in the point spread function measurement, the resulting volume was convolved with a uniformly labelled 170 nm sphere. 

Because no noise was added to the simulated MFM image used in Figure 3 of the main text, reconstruction quality continues to improve for 12,000 iterations with $\lambda=0.0001$. The resulting reconstruction is in excellent agreement with the reference volume used to generate the simulated MFM image.

The 50 frames of MFM snapshot data used in the tracking plots of Figure 4 (main text) and the animation in Movie S1 were processed into 3D reconstructions with 1000 iterations using $\lambda=0.01$, as in the snapshot used for Figure 2.  By synchronized shuttering of the excitation source, frames were exposed for 20 ms out of the 40 ms available to reduce the effects of motion blur.

\clearpage

\clearpage

\clearpage

\clearpage

%\clearpage

\clearpage

%\clearpage

\clearpage

{
\noindent
\includegraphics[width=1.0\linewidth]{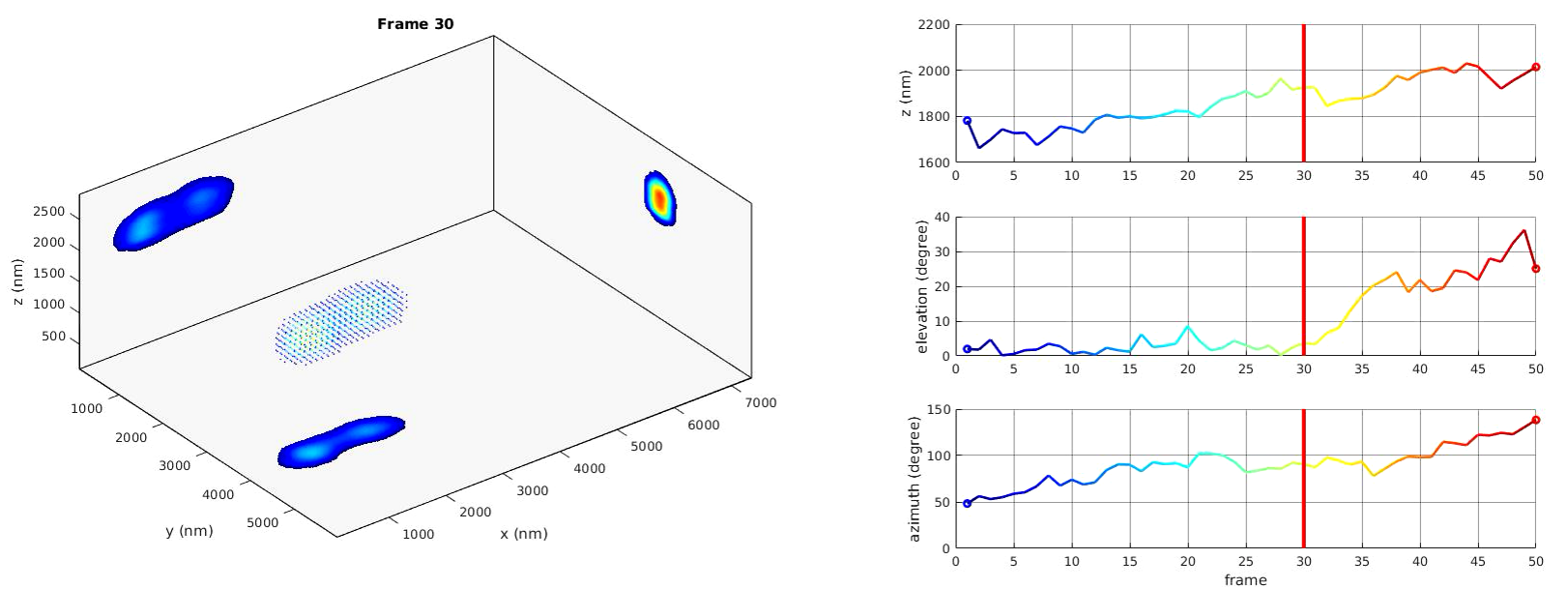}
}

\noindent
{\bf Caption for Movie S1}: A 3D animation of a single, fixed organism (\textit{P. fluorescens}) in water.  MFM snapshot images measured at 25 fps were analyzed for 3D distribution of fluorescence frame-by-frame.  The left half of the animation shows the bacterium rendered as cloud of points, one point per voxel colored by the intensity of the reconstructed emission from that voxel.  The organism is approximately 2 $\mu$m $\times$ 0.5 $\mu$m $\times$ 0.5 $\mu$m in size, and has enhanced fluorescent concentrations in two lobes along its major axis (i.e. at the poles). The $z$ position of its centroid is plotted at the top right.  The elevation angle of the major axis measured from the $x-y$ plane is plotted below that. % And plotted at the bottom right is the 
The azimuth angle of the major axis projected onto the $x-y$ plane, measured from the $y$-axis is plotted at the bottom right.  At the outset, the organism is horizontal and appears to be near the bottom cover slip which would keep it from tilting in elevation angle. When it rises high enough in $z$, around frame 30, it can then tilt more freely in elevation.

\clearpage

\bibliography{snapshot}

\section*{Acknowledgments}
We acknowledge helpful discussions with Ben Glick and Robin Graham.
This work was supported by funding through the Biological Systems Science Division, Office of Biological and Environmental Research, Office of Science, U.S. Department of Energy, under Contract DE-AC02-06CH11357.

\section*{Contributions}
XH, MH, NJF, AS, MKD, NFS, RW, AKK and OC contributed to writing the manuscript.
AS, XW, MKD, and NFS designed and implemented the microscope.
AS and XW performed the laboratory experiments.
RW developed the engineered biological organisms.
XH, MH, NJF, SY, KH, AKK, and OC developed the computational imaging approach and software.
XH, MH, NJF, KH, and SY performed the computational experiments.

\section*{Additional information} 
The authors have no competing interests to declare.

\vspace{0.5in}
\paragraph*{COPYRIGHT NOTICE:} This work was supported by funding through the Biological Systems Science Division, Office of Biological and Environmental Research, Office of Science, U.S. Dept. of Energy, under Contract DE-AC02-06CH11357.
The submitted manuscript has been created by UChicago Argonne, LLC, Operator of Argonne National Laboratory ('Argonne'). 
The U.S. Government retains for itself, and others acting on its behalf, a paid-up nonexclusive, irrevocable worldwide license in said article to reproduce, prepare derivative works, distribute copies to the public, and perform publicly and display publicly, by or on behalf of the Government.
The Department of Energy will provide public access to these results of federally sponsored research in accordance with the DOE Public Access Plan. http://energy.gov/downloads/doe-public-access-plan.

\end{document}